\documentclass{PoS}

\newcommand{\beq}{\begin{equation}}
\newcommand{\eeq}{\end{equation}}
\newcommand{\bea}{\begin{eqnarray}}
\newcommand{\eea}{\end{eqnarray}}

\usepackage{epsfig}
\input epsf

\title{Conformality at large number of fermion flavors and composite Higgs}

\ShortTitle{Conformality and composite Higgs}

\author{\speaker{E. T. Tomboulis}
\\
        Dept. of Physics and Astronomy, University of California, Los Angeles\\
        Los Angeles, CA 90095, USA\\
        E-mail: \email{tomboulis@physics.ucla.edu}}

\abstract{The many open questions concerning the existence of IR and UV fixed points in gauge theories as a function of the number of fermion flavors and bare coupling are briefly reviewed and discussed. It is pointed out that only a small subset of potential IR-conformal gauge theories, i.e. theories whose IR behavior is determined by an IR fixed point, has so far been examined. It is suggested that the naturally light scalar composites  
that seem to appear generically as lowest states in the non-QCD-like spectrum of such theories provide a natural basis for composite Higgs models, where the composite Higgs is not a NG boson 
and some of the usual fine-tuning problems are evaded. }  

\FullConference{31st International Symposium on Lattice Field Theory - LATTICE 2013\\
		July 29 - August 3, 2013\\
		Mainz, Germany}

\begin{document}

\section{Introduction}
Over the last several years a lot of effort has been devoted to exploring the possible existence and structure of IR fixed points (FP) in gauge theories with varying fermion flavor content and color representation \cite{DelD1}. 
Apart from their intrinsic QFT interest, a major motivation for these studies is potential application to  physics beyond the Standard Model (BSM). One such proposal is that of walking TC, where the number of flavors is such as to place a system just below the lower end of a conformal window. Many other possibilities, however, exist for BSM physics involving non-trivial IR FP's. 
In fact, one should note that, so far, only a small subset of possible IR-conformal gauge theories, i.e. theories whose long distance behavior is governed by an IR FP, has been explored. 

Here we will first briefly discuss what is known and the many open issues concerning the phase diagram, existence of FT's and the spectrum of states as a function of the number of flavors and the gauge coupling. On the basis of this discussion we outline how 
the naturally light (massless) scalar composites states appearing as lowest states in the spectrum of IR-conformal theories can lead to a wide class of models with composite Higgs which is not a NG boson.

\section{Phase diagram in $N_f$, $g$}
\begin{figure}[ht]
%\begin{center}
\includegraphics[width=6cm]{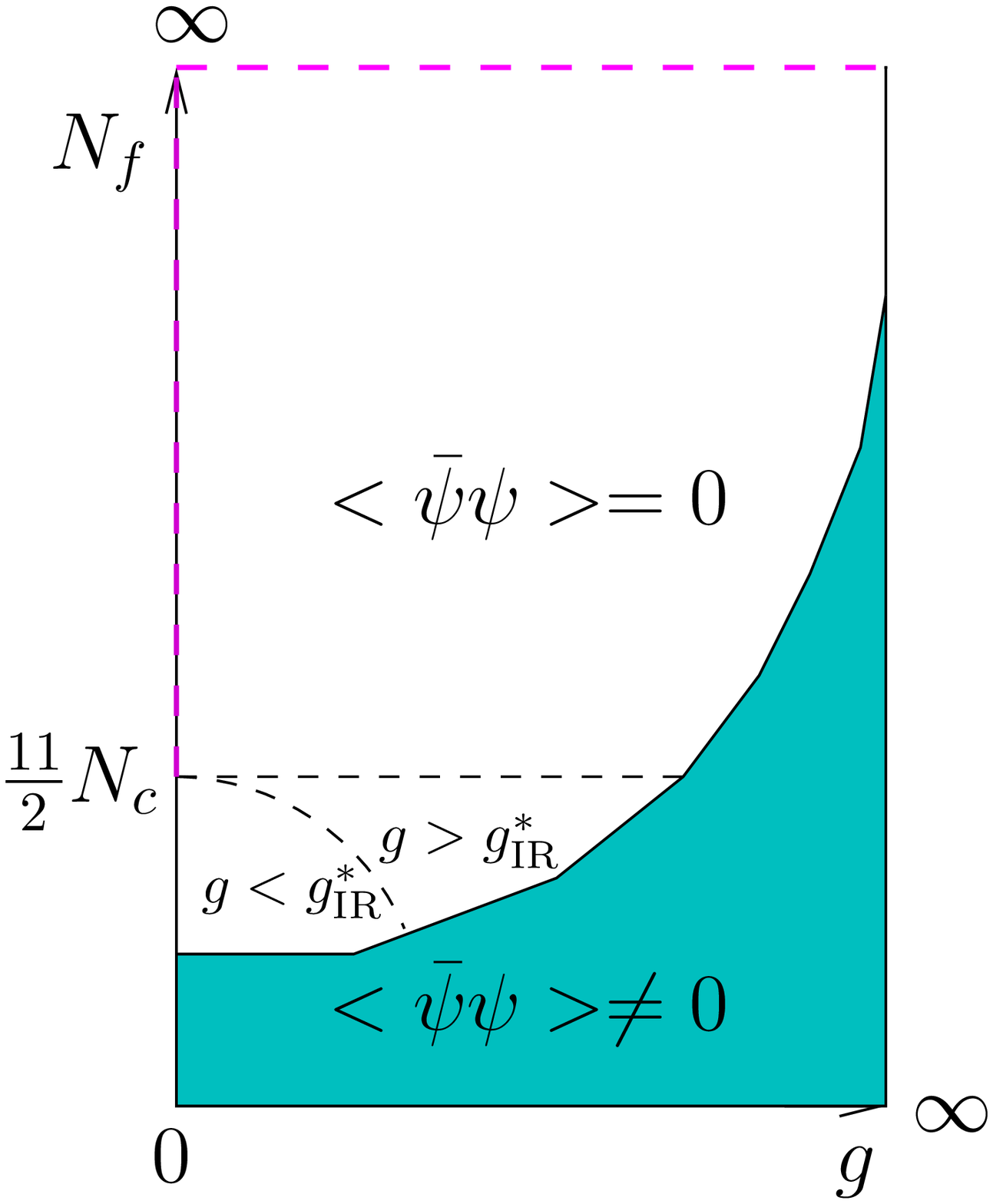}\hfill
\includegraphics[width=7cm]{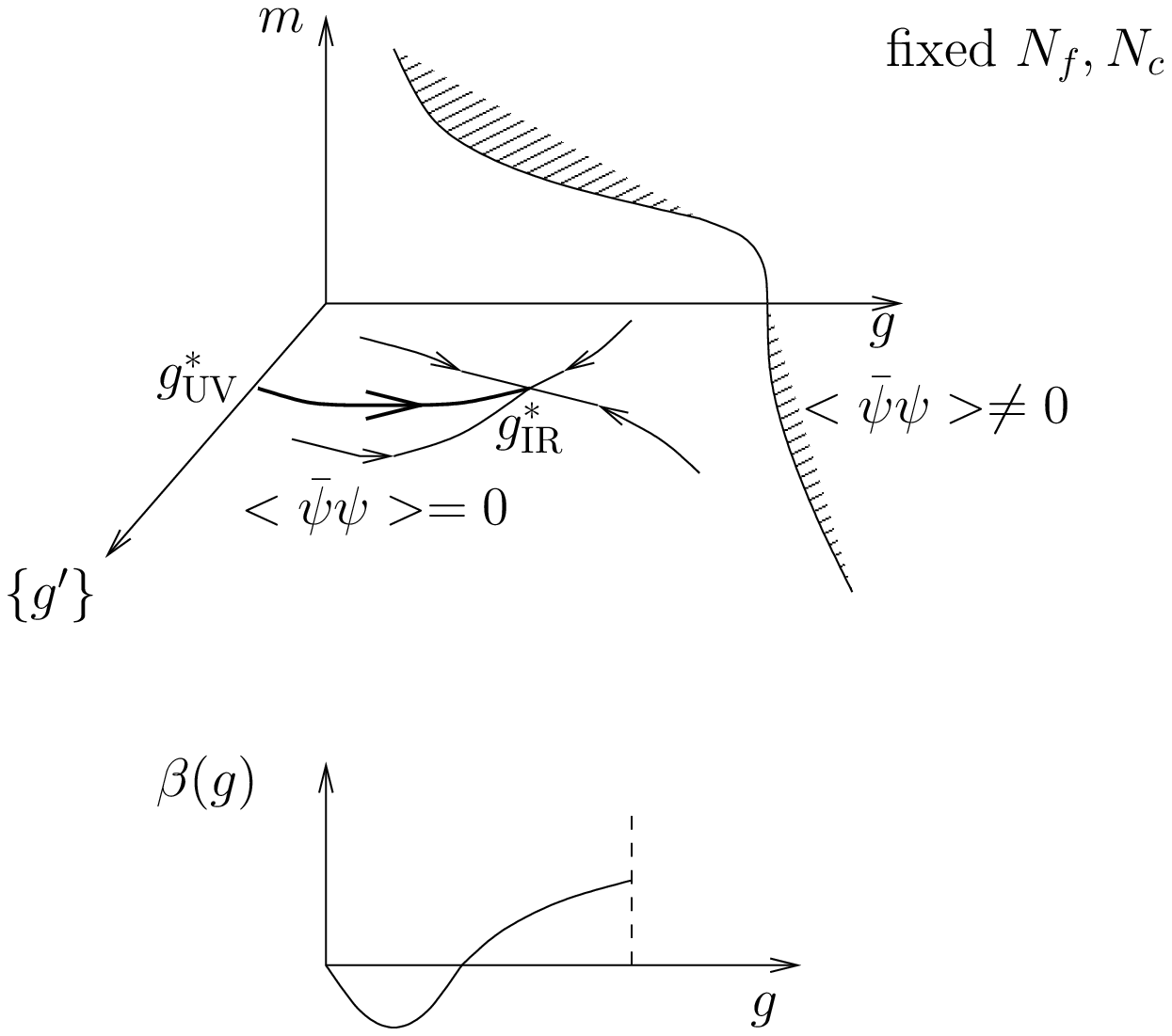}  
%\end{center}
\caption{Left: Phase diagram of $N_f$ vs. bare coupling $g$. Right: Standard picture within the putative CW ($N_f <  N_F^*$). $\{g^\prime\}$ denotes the set of irrelevant couplings and mass $m$ is the relevant direction.
\label{Fig1}
 }
\end{figure}

Recall that for a theory with $N_f$ fermion flavors in representation $R_f$ of a simple color group $G$ the perturbative 2-loop beta function possesses a non-trivial zero for a range of number of flavors 
$N_f^{**} < N_f < N^*_f$ defining a ``conformal window" (CW). Here $N^*_f= {11\over 4\kappa} {C_2(G)\over T(R_f)}$ where  
$\kappa=1(1/2)$ for 4 (2 ) - component fermions. For $G=SU(N_c)$ with fundamental rep.  fermions $N^*=11N_c/2\kappa$. Then, if $(N^*_f - N_f) << 1$ this zero is within the perturbative validity regime, and its existence can be trusted (Banks-Zaks IR FT). The perturbative value of the lower end $N_f^{**}$ 
of this range, however, cannot by trusted.  
Determining its actual (non-perturbative) value, i.e. the true extent of the CW, 
is a question that has been intensively investigated in recent years for a variety of fermion representations $R_F$ and mostly $G=SU(3)$ or $SU(2)$ \cite{DelD1}. 
In this connection, it has been commonly assumed that, for any number of fermion flavors,  chiral symmetry will eventually be broken provided the coupling is taken strong enough. 
This was recently found to be in fact incorrect.  
MC simulations for $N_c=3$ at inverse gauge coupling $\beta$ vanishing or small showed that chiral symmetry is restored via a first-order transition above a critical number of flavors ($\sim 52$ continuum) \cite{dFKU}. The same result was arrived at by resummation of the hopping expansion in the strong coupling limit: the familiar CSB solution abruptly disappears above a critical $N_f/N_c$ \cite{T1}.  Putting the available information together suggests the phase diagram of $N_f$ versus bare coupling $g$ at fixed $N_c$ in Fig.\ref{Fig1}-left below,  
shown for fundamental rep. $SU(N_c)$ fermions for definiteness, other cases being qualitatively similar. 
%[Along the vertical axis in the vicinity of bare coupling $g=0$ perturbation theory should hold, ]%[whereas in the limit $N_f \to \infty$ the theory becomes trivial. ] 

\begin{figure}[ht]
%\begin{center}
\includegraphics[width=6cm, height=5.5cm]{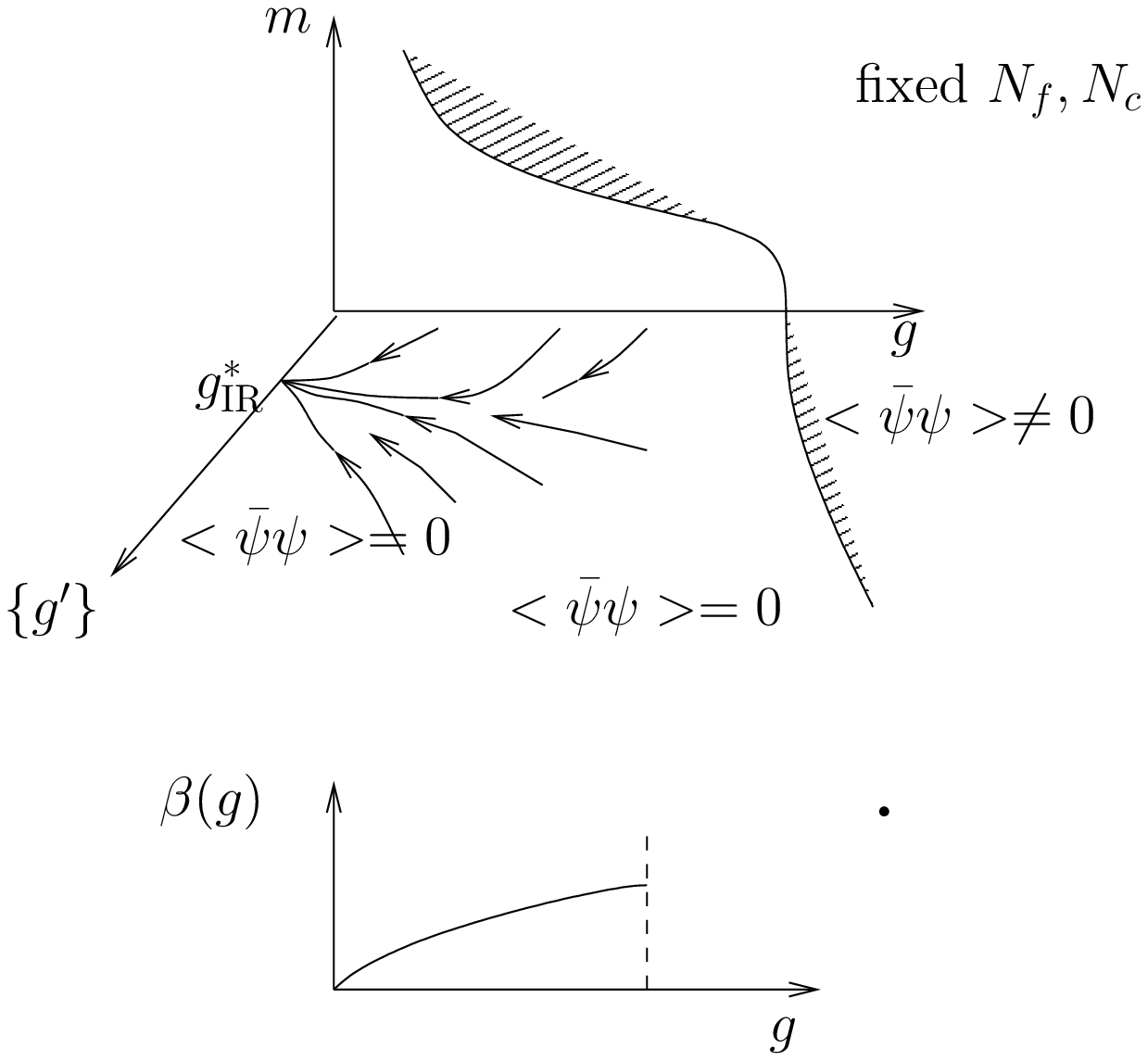} \hfill 
\includegraphics[width=6cm]{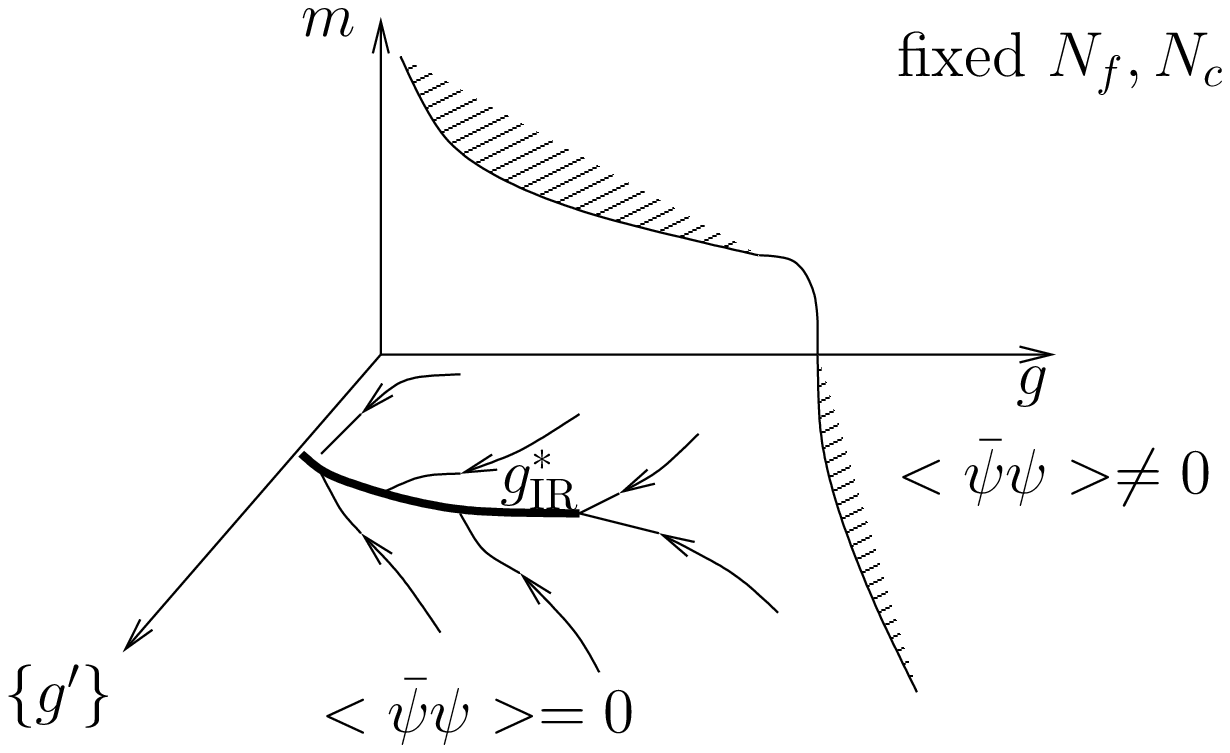}
%\includegraphics[width=6cm]{lat13fig8.eps}
%\end{center}
\caption{ Left: Lone trivial IR FT. Right: Line of IR fixed points with trivial end-point. Here $N_f > N_F^*$.}\label{Fig2}
\end{figure}

The boundary separating the chirally broken phase from the chirally symmetric phase terminates 
at infinite coupling at a finite critical $N_f$ as just explained. At the other extreme at vanishing coupling it terminates at a critical $N_f^{**}$ ($\sim 12$ for $N_c=3$) whose exact value remains 
controversial. The region between this boundary and the horizontal broken line 
is then the putative CW. The boundary is known to be a first order phase transition at least for some range starting from the strong coupling limit end ($g\to \infty$). Also, consistent with Fig. \ref{Fig1}-left, several studies over the years \cite{Misc} at fixed $N_f$ inside the CW have observed a first order phase transition to a chirally broken phase as the coupling is increased.\footnote{Additional, likely spurious, transitions may appear though at intermediate coupling depending on the particular fermion lattice action being used.} The simplest scenario would then be 
that the transition line is first order everywhere, though this might in fact depend on the 
fermion representation and gauge group. (Incidentally, a first order transition would spell trouble for standard walking TC.)  The standard picture of the RG flows  in the plane of irrelevant couplings inside the CW is shown in Fig.\ref{Fig1}-right. 
%Note that there is no physical distinction (any transition) between $g< g^*_{\rm IR}$ and $g>g^*_%{\rm IR}$ regions. 
The non-trivial IR FT $g_{\rm IR}^*$ inside the CW moves and merges with the UV FT at the origin as $N_f\to N_f^*$ from below, and 
perturbation theory (PT) for $N_f> N_f^*$ gives a trivial IR fixed point at $g=0$. 
The simplest scenario then is that this picture holds for all couplings (Fig. \ref{Fig2}-left). 

\begin{figure}[ht]
%\begin{center}
\includegraphics[width=6cm]{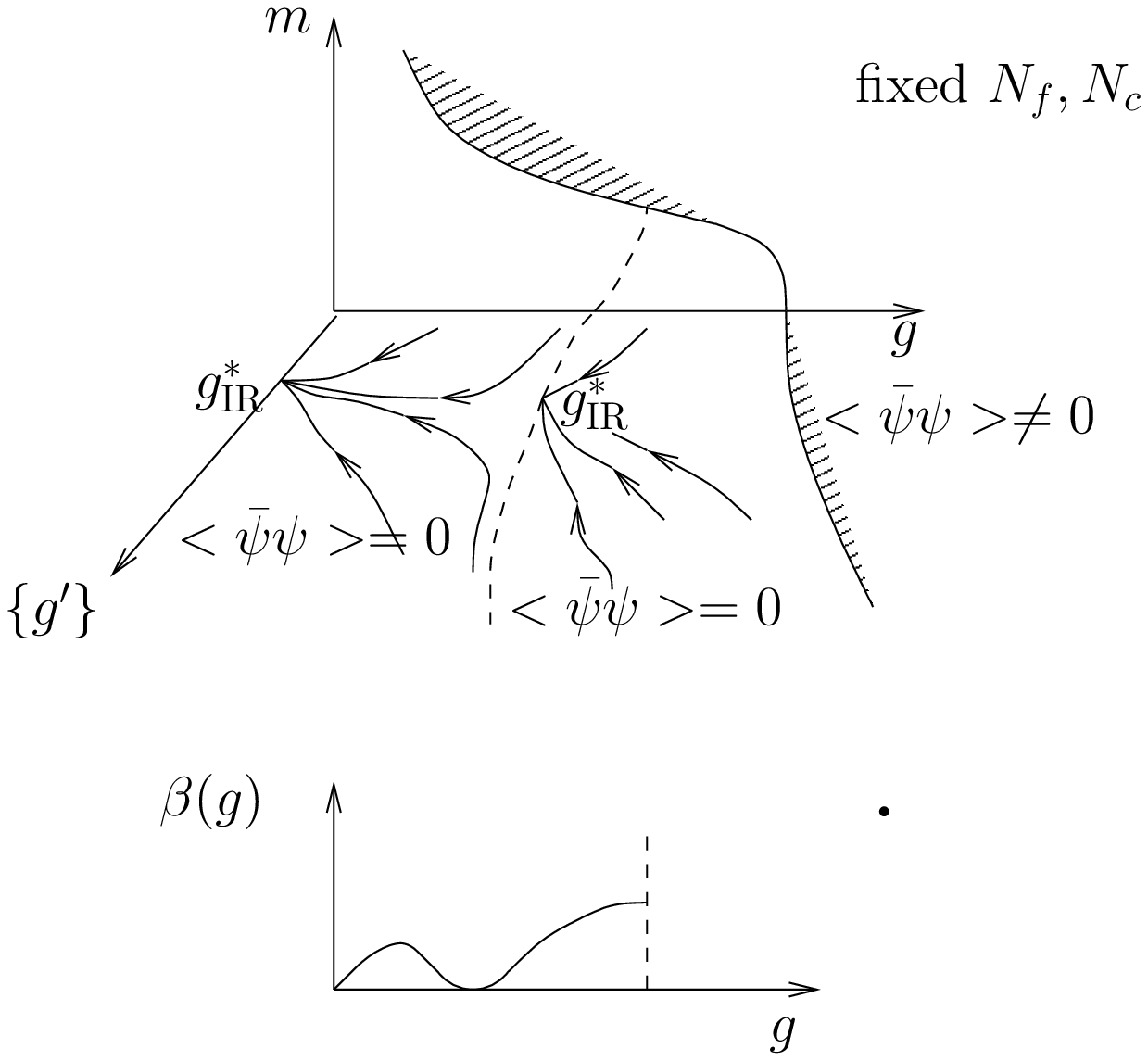}\hfill
\includegraphics[width=6cm]{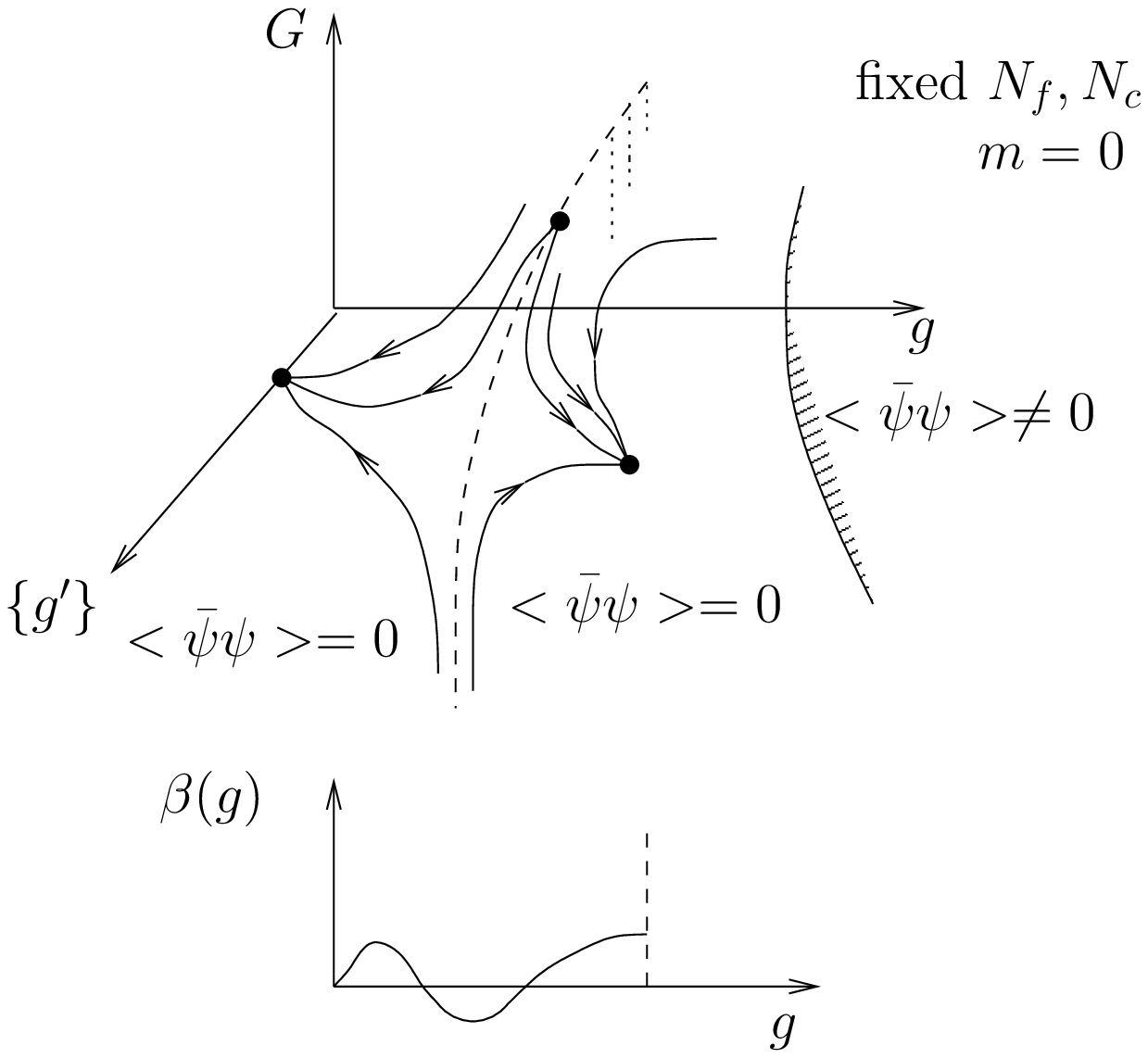}
%\includegraphics[width=6cm]{lat13fig7a.eps}
%\end{center}
\caption{Left: Non-trivial IR FT corresponding to an even beta-function zero. Right: Additional relevant direction $G$ resulting in nontrivial UV and IR fixed points.   Here $N_f > N_F^*$.} \label{Fig3}
\end{figure}

Life, however, could be rather more interesting. Simulation studies in \cite{dFKU} measuring toleron mass, Dirac spectrum and hadron spectrum for $N_c=3$ at zero or small $\beta$ find evidence for a nontrivial IR FP in the region  $N_f > N_f^*$. In view of these results it was conjectured in \cite{dFKU} that the FP location varies continuously with $\beta$, as well as $N_f$, reaching the value zero for $\beta\to \infty$, $N_f > N_f^*$, and for $N_f\to \infty$.  This amounts to a line of IR fixed points as depicted in Fig. \ref{Fig2}-right. 
This, however, would seem to contradict weak coupling PT where no such FP line ending at $g=0$ is seen. 
If such non-trivial IR FP's actually exist, their existence can be reconciled with PT if an even  
zero of the beta function obtains as depicted in Fig. \ref{Fig3}-left. Actually, one would expect such as zero to be unstable under changes in $N_f$ or other parameters unless perhaps it is an infinite order zero.  %(cf. below). 
More generally, though, it could appear as the limiting case of the situation shown in Fig. \ref{Fig3}-right. Here the possibility of other relevant directions (in addition to mass) is considered. 
These can arise from operators, such as chirally symmetric  4-fermi interactions, e.g., $G(\bar{\psi}\gamma_\mu\psi)^2$, whose anomalous dimensions at some intermediate couplings are such that they become relevant (marginal).  Massless quenched $\rm {QED}_4$ provides an example \cite{LLB}. 
As $N_f$ or other parameters are varied the non-trivial IR FT eventually merges with the non-trivial UV FT 
leading to the even degree zero in Fig. \ref{Fig3}-left,  which upon further increase of $N_f$ goes over to Fig. \ref{Fig2}-left consistent with the fact \cite{dFKU}, \cite{T1} that for $N_f\to \infty$ at fixed $N_c$ the theory becomes trivial. The situation depicted in Fig. \ref{Fig3} provides a possible scenario for reconciling the FT's found in \cite{dFKU}, coming from the strong coupling side, with weak coupling PT. It is, however, not the only one (cf. below). 

The situation depicted in Fig. \ref{Fig3} could actually arise also {\it within} the CW but, of course, with the sign of the beta-function reversed. This was discussed in \cite{KLSS}. It would amount to the occurrence of an UV FT beyond the non-trivial IR FT depicted in Fig. \ref{Fig1}-right. 
It should be emphasized in this connection that existing simulations exploring IR conformal behavior are all at fixed low $N_c$, typically 2,3.  They do not explore the regime of both $N_f, N_c$ large, 
as, for example, when they are adjusted so that a BZ-like IR FT occurs at large $N_f$. It is precisely this regime that is of interest for these new possibilities. Some indications already appear in considering the zeros of the beta-function 
in PT.  A general exploration of the non-trivial IR and UV zeros of the 4-loop beta function for a variety  of theories and representations is given in \cite{PS}. As always the question is whether  
such zeros can persist in the full theory. Typically, neglected terms in the beta function expansion when evaluated at the location of such zeros can be as large as the retained terms even when this location $\sim N_f^{-1}$ for large $N_f$.
Thus the unknown higher order corrections cannot be a priory neglected, and so, in contrast to the BZ FT, the existence of such zeros, even though perhaps suggestive, is not assured. 

There are also computations of the beta-function beyond standard PT, in particular within the $1/N_f$ expansion at fixed 'tHooft coupling $N_fg^2$. Because of the resulting non-local nature of vertices and inverse propagators in this expansion such computations cannot be performed exactly beyond leading order, but 
have been pursued in higher orders to a considerable degree within re-expansion schemes \cite{H}. 
A rich structure of IR and UV FT's in distinct branches of the beta-function in different regimes of the 'tHooft coupling is found, reminiscent of the supersymmetric pure $SU(N_c)$ beta-function 
\cite{KS}.  
In particular, for $SU(N_c)$, branches are found that can provide an alternative explanation for the IR FT's found in \cite{dFKU}. Again, however, it is hard to assess the reliability of these finding for the exact theory as the location and nature of the singularities (poles) delineating branches can be 
drastically altered by the higher omitted contributions \cite{H}, see also \cite{PS}.    

It is worth noting that all cases investigated so far form a small subset of possible IR-conformal gauge theories. In particular, only simple color groups have been considered. 
If the color group is semi-simple, e.g.,  $SU(N_1)\times SU(N_2)\times \cdots \times SU(N_n)$,  
there are $n$ gauge couplings resulting in a coupled set of equations for their beta functions.  
There are now correspondingly many choices for the coupling of fermions: different fermion subsets may be coupled to different subsets of group factors in different representations.  Depending on the number of such parameters available, many more possibilities for non-trivial IR and UV FT's may now arise. In particular, any such fixed points at (relatively) weak couplings could be very important for elw phenomenology.

\section{Spectrum}
Calculations of the spectrum in theories believed to be inside the CW (cf. Fig. \ref{Fig1}) indicate the appearance of some general features \cite{Sp}. 
 
Relevant parameters away from conformality are a 
quark mass $\hat{m}=m/\mu=am$, and the lattice size $L$.  
%Coupling to other gauge interactions may of course provide explicit breaking 
Below a "locking" scale $M_l$ (which is specific theory dependent) a scaling regime obtains where physical mass ratios remain essentially constant. 
Hadron masses scale as : \quad $M_{\rm H} \sim \mu \hat{m}^{1/(1+ \gamma_m^*)}$ \cite{DelDZ}. 
The detailed ordering of the mass spectrum is theory dependent but generally non-QCD like. 
Striking features are that the 
$0^{++}$ states are lowest and, also, that  gluonic states are below the lowest meson (scalar, pseudoscalar, vector) states. This picture, already suggested from analytic considerations around a BZ FT in \cite{M1}, is supported by subsequent lattice simulations \cite{Sp}.

For  $m\to 0 $ (or some other relevant deformation parameter such as $L\to \infty$)  the spectrum collapses to only massless states (``unparticles"). 
But at any small non-zero deformation such as a non-vanishing $m$ or finite box size one has a particle spectrum, with mass gap as 
above, containing a light scalar $0^{++}$ meson and a scalar gluball $0^{++}$ state plus the (somewhat heavier) rest of the  glueball and meson/baryon spectrum.
%\[ \sim \bar{\psi}\psi, \quad \bar{\psi}\gamma_5\psi, \quad \bar{\psi}\gamma_k\psi, 
%\quad \cdots  \]

The presence of naturally light scalar states suggests the following application. 

\section{Composite Higgs in IR conformal theories} 

Consider a  theory with $N_f$ `techniflavors' and  $N_c$ `technicolors' such that its IR behavior is controlled by an IR FP.   This IR FT may arise from any of the situations reviewed above, either inside the so-called CW or above it. For phenomenological reasons we would prefer it to be at weak coupling. Note, however, that, as seen from our previous discussion, this does not necessarily mean that the formation of the composites in the spectrum originated in a weak coupling regime. As we also saw, this generally implies that $N_f$, and correspondingly $N_c$, must be adjusted to be large. In fact, it is most natural to consider a non-simple color group, of at least two factors (accommodating a dark sector - see below), which opens an even wider set of possibilities for obtaining such an IR FT. The basic idea now is the following. 

At small deformation (e.g., a small quark mass $m$ or large finite box size $L$) one has a well-defined discrete spectrum of 
(weakly coupled) light scalar and other composite states. Coupling now other gauge interactions, 
in particular elw interactions, renders this system of (arbitrarily) light states unstable under the Coleman-Weinberg  mechanism. The resulting mass gap is now in effect the dynamically generated 
conformality deformation, which persists in the limit where the original explicit deformation is removed  ($m\to 0 $ or $L\to \infty$).  
This coupling of elw interactions directly to the naturally light (massless) states present in an IR-conformal theory allows for  a wide class of potential composite Higgs models, distinct 
from theories of the Higgs as a NG boson or walking TC.

As an example of this type of model, consider an IR-conformal theory 
%$SU(N)$ (or $U(N)$) 
with $N_f$ flavors in the fundamental rep. of the `technicolor` gauge group.  
%such that system in CS phase. 
Single out just two of these flavors $Q=(U,D)$ (to be later coupled to elw $SU(2)\times U(1)$). 
There are now composite  states formed by $Q$ and the remaining fermion flavors $\psi_a$, $a=1, \ldots, 
N_f-2$, such as $ \bar{\psi}\psi$, $\bar{\psi}Q$, $\bar{Q}Q,  \ldots$. 
The ``mixed" sector can be eliminated by taking semi-simple color gauge group, 
e.g., $SU(N_1)\times SU(N_2)$ with $\psi$ charged under both factors, and the $Q$ charged under only one factor. 
 `Mixed' composites such as $\bar{\psi}Q$, $\psi QQ, \ ...$ no longer form. The only possible color singlet mixed states that could form are highly unstable multi-quark (tetra and higher) states if they form at all.  

Now consider coupling the elw interactions. This may of course be done in various ways depending how they are to be aligned relative to our IR-conformal theory. Here we just couple to the singled-out $Q$ fermions as follows.  
The scalar meson $\bar{Q}Q$ gives rise to the four fields: \\ 
%$h_i, i=0, 1, 2,3$: 
$ h^+ = - \bar{D}U,  \quad h^-= \bar{U}D, \quad h_3= (\bar{U}U - \bar{D}D)/\sqrt{2}, 
\quad  h_0= (\bar{U}U + \bar{D}D)/\sqrt{2}  $.  
These may be taken to form the weak scalar doublet 
\beq \quad H= \left(\begin{array}{c} h^+ \\ (h_0+ ih_3)/\sqrt{2}\end{array} \right) \; , \qquad 
\tilde{H}= i\tau_2 H^*= \left(\begin{array}{c} (h_0- ih_3)/\sqrt{2} \\ h^- \end{array}\right) 
\eeq  
after giving $Q$ ordinary quark elw charges. 
In addition one, of course, has the other pseudoscalar $P= \bar{Q}\gamma_5 Q$, 
vector $V_k= \bar{Q}\gamma_k Q$, etc. meson states, as well as baryon states, and the  `dark' 
sector scalar $\Psi$ and other meson and baryon states formed by the remaining flavors $\psi_a$. 
The glueball states are all weak singlets (completely dark). In particular one has the $0^{++}$ 
state, which,  together with the $h_0$, are expected to be the lightest states. These two lightest scalar states may in general mix.  
%so one expects two independent (mixtures) scalar states. 
Note that the elw interactions only couple to the fermionic component. 

The effective theory of the IR-conformal `techni'-interactions at low energies is in principle obtained by matching the composites to interpolating fields. Schematically, this will result in an effective potential of the form 
\bea 
 \lambda (H^\dagger H)^2 + \lambda_1(P^\dagger P)^2 + \lambda_2 |P^\dagger H|^2 + \cdots 
\lambda_V |V^\dagger_kV_k|^2 
%& &  \nonumber \\ 
+ \cdots  + \lambda_d (\Psi^\dagger\Psi)^2  + 
  \lambda_d^\prime (\Psi^\dagger\Psi)(H^\dagger H) + \cdots  \label{effpot}
  %& & 
\eea  
Assuming that the  IR FP occurs at (relatively) weak coupling, all effective couplings in this long distance effective potential are weak. 
The coupling to the electroweak gauge fields renders this system of (nearly) massless scalar fields 
unstable under the Coleman-Weinberg mechanism. 
With parity and Lorentz symmetry assumed preserved, only the scalar  $H$ condense - which results in the usual elw breaking pattern. From the available deformed spectra computations it is not hard to envision that a gap of a factor of, say, 5-7 can develop between the lightest massive scalar (physical Higgs) and the higher massive states resulting from this breaking. 
One may, of course, analogously introduce other gauge interactions among the $\psi_a$'s so that, through Coleman-Weinberg or other mechanisms, the dark sector is rendered massive or it consists of massive and unparticle sectors. 
%Conformal dilaton mechanism still applicable in this context. 

SM quark masses could, as usual, be introduced by the  extended  TC mechanism of effective 4-fermi interactions  between Q fermions and the SM quarks  at a high scale. The interpolating field 
matching to the techni-composites converts such interactions to effective Yukawa couplings for the SM quarks. Note that no techniquark condensate at some intermediate scale, with its usual attending fine-tuning problems, is here involved.  

In summary, it should be clear that a variety of new composite Higgs models could be devised along these lines. The real difficulty is identifying the FT structure and spectrum of a candidate IR-conformal theory and extracting quantitative information from it as needed for model building in each particular instance.   

\vspace{0.2cm}
The author thanks the Aspen Center for Physics where part of this work was done and the participants of the workshop "LGT at the LHC era" (June 2013) for many discussions.

\end{document}